\documentclass[12pt]{iopart}
\usepackage{iopams}
\usepackage{amssymb,bm,graphicx}
\usepackage{color}

\newcommand{\ii}{\mathrm{i}} 
\newcommand{\tfrac}{\frac}

\newcommand{\figpath}{}

\begin{document}

\title[Singularity formation in thin-film flow equations]{Self-similar finite-time singularity formation in degenerate parabolic equations arising in thin-film flows}

\author{M~C~Dallaston$^1$, D~Tseluiko$^2$, Z~Zheng$^3$, M~A~Fontelos$^4$, and S~Kalliadasis$^1$}

\address{$^1$ Department of Chemical Engineering, Imperial College London, London SW7 2AZ, UK}
\address{$^2$Department of Mathematical Sciences, Loughborough University, Loughborough LE11 3TU, UK}
\address{$^3$Department of Mechanical and Aerospace Engineering, Princeton University, Princeton New Jersey 08544, USA}
\address{$^4$Instituto de Ciencias Matem\'aticas (ICMAT), C/Nicol\'as Cabrera, Madrid 28049, Spain}

\submitto{\NL}

\begin{abstract}
A thin liquid film coating a planar horizontal substrate may be unstable to
perturbations in the film  thickness due to unfavourable intermolecular
interactions between the liquid and the substrate, which may lead to
finite-time rupture. The self-similar nature of the rupture has been
studied before by utilizing the standard lubrication approximation along
with the Derjaguin (or disjoining) pressure formalism used to account for
the intermolecular interactions, and a particular form of the disjoining
pressure with exponent $n=3$ has been used, namely, $\Pi(h)\propto
-1/h^{3}$, where $h$ is the film thickness. In the present study, we use a
numerical continuation method to compute discrete solutions to self-similar
 rupture for a general disjoining pressure exponent $n$.
 We focus on axisymmetric point-rupture solutions and show that pairs
of solution branches merge as $n$ decreases, leading to a critical value
$n_c \approx 1.485$ below which stable similarity solutions do not appear
to exist. We verify that this observation also holds true for plane-symmetric line-rupture solutions for which the critical value turns out to be slightly larger than for the axisymmetric case, $n_c^{\mathrm{plane}}\approx 1.5$. Computation of the full time-dependent problem also demonstrates
the loss of stable similarity solutions and the subsequent onset of cascading oscillatory structures.
\end{abstract}

\section{Introduction}

Thin liquid films are ubiquitous in a wide spectrum of natural phenomena and
technological
applications~\cite{Craster2009,Grotberg2004,Kalliadasis2012,Sharma1985}. For
sufficiently small film thicknesses, intermolecular forces can be of the same
order as the classical hydrodynamic effects such as viscosity and surface
tension~\cite{Bonn2009,DeGennes1985,DeGennes2013,Oron1997}. For a film on
a solid substrate, these intermolecular forces can cause rupture of the film
and subsequent dewetting of the
substrate\,\cite{Bertozzi2001,Witelski1999,Witelski2000,Witelski2004,Zhang1999}
and they also play a major role in contact line dynamics in
general\,\cite{Hocking1994,Saprykin2007,Schwartz1998}. For a free liquid film
or jet, intermolecular forces can also cause
rupture/breakup\,\cite{Erneux1993,Papageorgiou1995,Vaynblat2001}.
Combinations of these two situations occur, for instance, with multiple
stacked fluid layers on a substrate\,\cite{Ward2011}.

Mathematical models of thin-film dynamics usually take advantage of
lubrication theory to reduce the Navier--Stokes equations and associated wall
and free-surface boundary conditions to an evolution equation for the film
thickness $h(\boldsymbol x,t)$, a function of space $\boldsymbol x$ and time
$t$ \,\cite{Oron1997}. The governing equation for $h(\boldsymbol x,t)$ has
the general form
\begin{equation}
\label{eq:genpde}
h_t = -\nabla\cdot(\mathcal M(h) \nabla \nabla^2 h) + \nabla\cdot(\mathcal N(h) \nabla h),
\end{equation}
where $\mathcal M$ and $\mathcal N$ are nonlinear functions of the thickness.
As well as modelling dewetting films ($\mathcal M(h)=h^3$, $\mathcal N(h)=-h^3 \Pi'(h)$, where $\Pi(h)$ is the disjoining pressure arising from intermolecular forces), equation (\ref{eq:genpde}) incorporates
nonlinear diffusion and the porous media equation ($\mathcal M(h) =0$)
\,\cite{Bertozzi1994a,Gilding1976,Peletier1981}, the flow of viscous gravity
currents ($\mathcal M(h) = 0$, $\mathcal N(h) = h^3$) \cite{Huppert1982}, capillary
dominated flows with Navier slip ($\mathcal M(h) = h^2(h+\lambda)$, $\mathcal
N(h)=0$)\, \cite{Greenspan1978,Hocking1981}, models of electrified thin
films (${\mathcal M(h) = h^3}$, ${\mathcal N(h) = -h^2 - \beta}$) \,
\cite{Conroy2010,Wang2011},  and models of heated thin films (${\mathcal M(h) = h^3}$, $\mathcal N(h) = h^2T'(h)$, where $T(h)$ represents the surface temperature of the film, e.g., Joo et al.\,\cite{Joo1991} and Thiele and Knobloch \cite{ThieleKnobloch2004} obtain $T(h)=1/(1+Bh)$, while Boos and Thess \cite{Boos1999}  model the surface temperature as a linear function of the film thickness, $T(h) = 1 - Bh$).  Mathematical properties of (\ref{eq:genpde})
for $\mathcal M(h) = h^n$ have been considered for general $n$, with
no second-order term ($\mathcal N(h) = 0$)\,\cite{Bertozzi1996,Bertozzi1994}, and
with a porous media-type term $\mathcal N(h) =
h^m$\,\cite{Bertozzi1994a,Bertozzi1998}.

\begin{figure}
\centering

\includegraphics{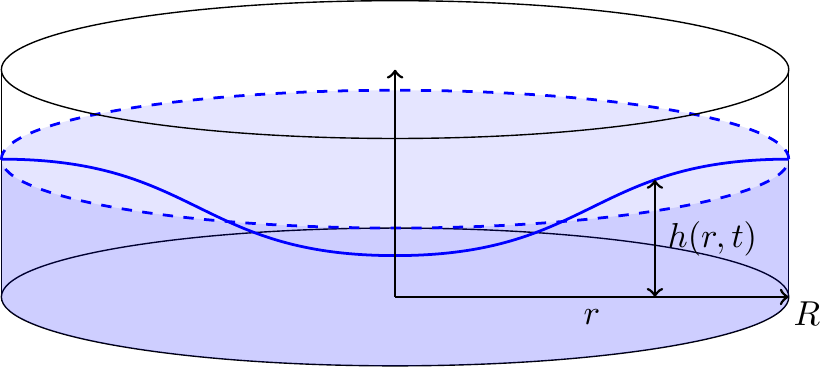}

\caption{A schematic of the axisymmetric thin-film problem (\ref{eq:pde}) in a finite domain $0<r<R$.}

\label{fig:schematic}
\end{figure}

In this study, we consider the effect of intermolecular forces on finite-time
point rupture of a dewetting film on a solid substrate.  The film is assumed
to be axisymmetric around a point, as is the case for a thin film contained
within a circular dish of radius $R$ (see figure \ref{fig:schematic}). The thickness of the
film $h(r, t)$ is modelled by the (dimensionless) lubrication equation
\begin{equation}
\label{eq:pde}
h_t = -\frac{1}{r}(r q) _r, \qquad
q = h^3\left( \frac{1}{r} (rh_r)_r + \Pi(h) \right)_r
\end{equation}
where  $r$ is the radial coordinate and $q$ is the flow rate in the radial direction per unit length in the
angular direction (see figure
\ref{fig:schematic}). Subscripts denote partial derivatives. The first term
in $q$ is the Laplace pressure (linearised curvature), while the disjoining
pressure is taken to be of the form
\begin{equation}
\Pi(h) = -\frac{1}{nh^n}, \qquad n > 0.
\label{eq:pi}
\end{equation}
Equation~(\ref{eq:pde}) expresses the balance between surface tension,
viscosity and intermolecular forces.
 We note that (\ref{eq:pde}) has been non-dimensionalised so that it is independent of the material parameters. To achieve this, we use a characteristic film thickness $h_0$ as the thickness scale, $\ell=(\gamma h_0^{n+1}/A)^{1/2}$ as the lateral lengthscale, where $\gamma$ is the liquid-gas surface tension coefficient and $A$ is the dimensional coefficient multiplying the dimensional disjoining pressure term (note that for the case of van der Waals interactions with $n=3$, $A=A^*/6\pi$, where $A^*$ is the so-called Hamaker constant), and $T=3\mu\gamma h_0^{2n-1}/A^2$ as the time scale, where $\mu$ is the dynamic viscosity of the liquid. The thickness scale $h_0$ can be chosen to be equal to 
 the undisturbed film thickness without loss of generality, so that the undisturbed dimensionless film thickness is unity.
 In a finite domain, the non-dimensional location of the outer boundary $r=R$ is then a parameter.

In applications, the appropriate value
of the exponent $n$ depends on the origin of the disjoining pressure, with,
for example, $n=1$ for hydrogen bonds, $n=2$ for surface
charge-dipole interactions, and $n=3$--$4$ for van der Waals
forces\,\cite{Nold2014, Nold2015,Teletzke1987,Teletzke1988}. Using elements
from the statistical mechanics of fluids, namely, density-functional theory
(DFT), which takes into account the non-local character of the intermolecular
interactions via an appropriate free-energy functional, Yatsyshin et
al.~\cite{Yatsyshin2015} showed that for a film on a planar wall, the local
disjoining pressure with $n=3$ is an asymptote to DFT as the distance of the
chemical potential from its saturation value vanishes. For general $n$,
(\ref{eq:pde}) is of the form (\ref{eq:genpde}) with $\mathcal M(h) = h^3$ and
$\mathcal N(h) = -h^{2-n}$.  Note that the special case $n=0$ ($\mathcal N(h) = -h^{2}$) is equivalent to a disjoining pressure $\Pi = \log(h)$, which arises in thermocapillary film flow.

With the sign in (\ref{eq:pi}), the
disjoining pressure acts like negative diffusion, destabilising a flat film
(in contrast to \cite{Bertozzi1994a}); while the negative diffusion equation
is ill-posed, Bertozzi and Pugh~\cite{Bertozzi1998} showed that the Laplace
pressure term prevents finite-time blow up of solutions, at least in one
spatial dimension.

Multiple stabilising or destabilising effects may be included in the
disjoining pressure, for instance in a two-term model \cite{Schwartz1998}.
However, here we consider the effect of a single, destabilising term. Of
course, boundary conditions for (\ref{eq:pde}) must be included to fully
specify the problem; we will discuss the appropriate conditions in sections
\ref{sec:similarity} and \ref{sec:numerics}.

There are examples of equations of the form (\ref{eq:genpde}) which feature
finite-time rupture at a time $t_0$, that is, $h \rightarrow 0$ at a point or
points as $t\rightarrow t_0 < \infty$, and it is of theoretical interest to
know which functions $\mathcal M$ and $\mathcal N$ allow this to
happen\,\cite{Bertozzi1994}. Self-similar rupture dynamics of (\ref{eq:pde})
have been considered in depth for $n=3$, appropriate for van der Waals forces
\cite{Witelski1999,Witelski2000,Zhang1999}.  In this case, it has been shown
that a discrete number of similarity solutions to (\ref{eq:sim}) can be
constructed that satisfy the appropriate stationary far-field condition.
These solutions were computed numerically, using a shooting
method\,\cite{Zhang1999}, and Newton iterations on a discretised
boundary-value problem\,\cite{Witelski1999}.  In each case, the numerical
computation is highly sensitive to the initial guess (the right-hand initial
condition for shooting, or the initial guess of the Newton scheme,
respectively).

Similarity solutions to the plane-symmetric version of (\ref{eq:sim}) have
also recently been computed\,\cite{Tseluiko2013} using the continuation
algorithms implemented in the open source software
AUTO07p\,\cite{Doedel2007}, where it was seen that numerical continuation
provides a convenient way of both choosing a starting point for computations,
and also constructing the discrete solutions. These are achieved by
introducing artificial `homotopy' parameters that are varied gradually. In
addition, the selection mechanism in the plane-symmetric version was explored
in \cite{Chapman2013}, where the exponential asymptotics in the large
branch-number limit was performed.

The aim of this work is to extend the study of the rupture dynamics of thin
films from $n=3$ to general exponent $n$. We perform computations of
similarity and time-dependent solutions to (\ref{eq:pde}).  As well as an
exploration of the mathematical properties of (\ref{eq:pde}), our results
may also shed light on the behaviour of films affected by other types of
intermolecular forces and additional complexities.

In section \ref{sec:similarity}, we extend the computation of discrete
similarity solutions of (\ref{eq:pde}) to general exponent $n$, as well as
show how the numerical continuation methods employed in \cite{Tseluiko2013}
for the plane-symmetric case may be extended to the axisymmetric version.  In
addition to providing a simple method to find the discrete solutions,
numerical continuation also allows us to trace out the discrete solution
branches as $n$ is varied. In section \ref{sec:numerics}, we solve the
time-dependent problem in a finite domain with appropriate boundary
conditions. Conclusions are offered in section \ref{sec:conclusion}.

\section{Self-similar rupture}
\label{sec:similarity}

\subsection{Formulation}
Assuming self-similarity near a rupture point ($r=0$), the
film thickness may be expressed as
\begin{equation}
h(r,t) = (t_0-t)^\alpha f(\xi),
\label{eq:similarity_ansatz}
\end{equation}
where $\xi$ is the similarity variable given by
\begin{equation}
\xi =\frac{r}{(t_0-t)^\beta}.
\end{equation}
Substituting (\ref{eq:similarity_ansatz}) in (\ref{eq:pde}), we find that $f$ satisfies the ordinary differential equation
\begin{equation}
-\alpha f + \beta\xi f' = -\frac{1}{\xi}\left[\xi f^3\left(f'' + \tfrac{1}{\xi}f' \right)' + \xi f^{2-n}f' \right]',
\label{eq:sim}
\end{equation}
where the similarity exponents $\alpha$ and $\beta$ are determined uniquely by matching the powers of $(t_0-t)$ premultiplying each of the viscous, capillary and disjoining pressure terms:
\begin{equation}
\alpha = \frac{1}{2 n-1}, \qquad \beta = \frac{ n+1}{4 n-2}.
\label{eq:alphabeta}
\end{equation}

The fourth-order similarity equation (\ref{eq:sim}) must be closed by four boundary conditions.  Requiring that $h$ be smooth at the origin $r=0$ when $t < t_0$ provides the left-hand conditions
\begin{equation}
\label{eq:simbc1}
f'(0) = f'''(0) =0.
\end{equation}
Conditions at the right-hand boundary are derived from the requirement of quasi-stationarity\,\cite{Zhang1999}.  For $n>1$, $h_t = (t_0-t)^{\alpha-1}(-\alpha f + \beta\xi f')$ is assumed to be unbounded at the rupture point as $t\rightarrow t_0$, while the velocity of the interface far from rupture should remain bounded.  The appropriate dominant balance in (\ref{eq:sim}) as $\xi\rightarrow\infty$ is, therefore, between the first two terms. Thus,
\begin{equation}
f \sim {c_0}\xi^{\alpha/\beta}, \qquad \xi \rightarrow\infty.
\label{eq:ff}
\end{equation}
For general $n$, ${c_0}\xi^{\alpha/\beta}$ is not an exact solution of equation (\ref{eq:sim}) (this, however, is the case for $n=3$, for which $\alpha=1/5$ and $\beta=2/5$). Nevertheless, a full asymptotic expansion (as $\xi\rightarrow \infty$) in decreasing powers of $\xi$, satisfying the similarity equation~(\ref{eq:sim}), can be found in the form
\begin{equation}
f(\xi) \sim F(\xi) = \sum_{k=0}^\infty F_k(\xi),
\label{eq:asympt_expansion}
\end{equation}
where $F_k(\xi)=c_k \xi^{\alpha/\beta-a_k}$ with $a_0=0$ and $a_k$, $k=1,2,\,\ldots,$ being an increasing sequence of positve numbers that depend on $n$, and $c_k$, $k=1,2,\,\ldots,$ being the coefficients that can be uniquely determined by the value of $c_0$.

In fact, the condition
\begin{equation}
f(\xi) \sim F(\xi), \qquad \xi \rightarrow\infty,
\label{eq:ff2}
\end{equation}
effectively imposes two boundary conditions on (\ref{eq:sim}). This may be deduced by analysing the `stability' of  $F(\xi)$ as $\xi\rightarrow\infty$ using the WKB ansatz
\begin{equation}
f \sim F + \epsilon \tilde f, \qquad
\quad \tilde f \sim \varphi(\xi) \exp(\lambda \xi^p), \qquad \xi \rightarrow \infty.
\end{equation}
Linearising (\ref{eq:sim}) for small $\epsilon$ results in
\begin{eqnarray}
\label{eq:fflinear}
-\alpha \tilde f + \beta \xi \tilde f' + \frac{1}{\xi}\Big[ \xi F^3(\tilde f'' + \tfrac{1}{\xi}\tilde f')' &+ 3\xi F^2(F'' + \tfrac{1}{\xi}F')\tilde f \\
 &\!\!\!\!\!+ F F^{2- n}\tilde f' + \xi(2- n)F^{1-n}F'\tilde f \Big]' = 0.
\end{eqnarray}
Note that only the leading-order behaviour in $\xi$ is required to determine the exponent $\lambda$.  Assuming $p > 1-\alpha/\beta$, which will be confirmed {\it a posteriori}, the dominant balance in (\ref{eq:fflinear}) is between the second term and the highest-order derivative in the expansion of the third term, that is,
\begin{equation}
\beta \xi \tilde f' \sim -F^3\tilde f''''.
\end{equation}
Substituting the assumed forms for $F$ (again, only the leading-order term $F_0$ is required) and $\tilde f$ and keeping the highest-order terms in $\xi$ results in
\begin{equation}
\beta \varphi p \lambda \xi^p = -{c_0}^3 \varphi p^4\lambda^4\xi^{4(p-1)+3\alpha/\beta}.
\end{equation}
Finally, matching the exponents gives the value for the power $p$ of the function in the exponent, while matching the coefficients gives a quartic equation for the eigenvalue~$\lambda$:
\begin{eqnarray}
&&p = 4(p-1)+3\alpha/\beta \quad \Rightarrow \quad p = \frac{4}{3} - \frac{\alpha}{\beta} = \frac{2}{3}\frac{2 n-1}{ n+1},\\
&&\beta \lambda p = -{c_0}^3\lambda^4p^4 \quad \Rightarrow \quad \lambda \in \left\{0, -\frac{\beta^{1/3}}{{c_0}p}, \frac{\beta^{1/3}}{2{c_0}p}(1 \pm \ii\sqrt 3) \right \}.
\end{eqnarray}
Here we have used (\ref{eq:alphabeta}) to express $\alpha$ and $\beta$ in terms of $n$.  Note that $p > 1-\alpha/\beta$ as required.  In addition, $p$ is positive for any $ n > 1/2$ and thus there are two values of $\lambda$ with positive real parts (the zero eigenvalue allows for perturbations in the parameter~$c_0$).  The constants on the eigenfunctions $\tilde f$ corresponding to these values must be zero to achieve (\ref{eq:ff}), thus two constraints are imposed on the problem by this far-field condition.    When $ n = 3$, we have $p=5/6$, which was derived previously\,\cite{Chapman2013,Zhang1999}.

There are some critical values of $n$ to take note of.  Care must be taken in applying the far-field boundary conditions when $n\leq1$, as the quasi-stationarity condition only implies (\ref{eq:ff}) when $n > 1$.  In our computations of self-similar solutions, we will not observe solution branches that extend below $n=1$, so this potential issue is not of concern.
In addition, when $n < 1/2$, $\alpha$ and $\beta$ are negative, and the similarity ansatz~(\ref{eq:similarity_ansatz}) does not represent rupture.
At $n = 1/2$, the similarity exponents $\alpha$ and $\beta$ in (\ref{eq:alphabeta}) are undefined.  At this critical value, the scaling no longer behaves as a power law, and instead of (\ref{eq:similarity_ansatz}) the appropriate ansatz to make is
\begin{equation}
h(r,t) = \e^{\alpha t} f(\xi), \qquad \xi = r\e^{-\beta t},
\end{equation}
where $\beta = 3\alpha/4$ and $\alpha$ is undetermined from the scaling (it is possible in this case that $\alpha$ is determined as a second-kind similarity solution~\cite{Barenblatt1996}, given an appropriate quasi-stationary far-field condition, but we do not pursue this issue further in the current study).

Following \cite{Zhang1999}, we enforce the far-field condition (\ref{eq:ff}) approximately by applying two mixed conditions that are satisfied by (\ref{eq:ff}) at $\xi = L \gg 1$, the simplest of which are
\begin{equation}
\label{eq:simbc2}
L f'(L) - \frac{\alpha}{\beta} f(L) = 0, \qquad L^2 f''(L) - \frac{\alpha}{\beta}\left(\frac{\alpha}{\beta}-1\right)f(L) = 0.
\end{equation}
The task is now to compute solutions to (\ref{eq:sim}), for a given $n$, that satisfy the
boundary conditions (\ref{eq:simbc1}) and (\ref{eq:simbc2}).  As mentioned in
the Introduction, this will be achieved using numerical continuation.

\subsection{Solutions by numerical continuation}

The general aim of numerical continuation is to compute a solution to a
nonlinear, algebraic system of equations (including discretisations of
differential equations), which feature one or more parameters. The dependence
of the solution on a given parameter, that is, the solution branch, is traced
out numerically using a predictor--corrector method: given one point on the
branch (for a given parameter value), a small step is taken along the
branch, varying both the parameter and the solution, to a new point on the
branch.  As this step can only be taken approximately, this predictor is then
used as the initial guess in a correction algorithm, usually a variation of
Newton's iteration.  This process may be repeated to compute the branch over
a longer interval. It is advantageous to parameterise the solution branch by
its arclength (or at least a local approximation), rather than the system
parameter, in order to easily handle turning points in the solution branch.
In this case, the method in known as pseudo-arclength continuation\,\cite{Allgower2003}.

The parameters used in a continuation procedure may be model parameters or
artificial parameters introduced for expediency, and we will make use of both
in the following. The computations are done with software
AUTO07p\,\cite{Doedel2007}, which utilises the pseudo-arclength continuation method and
has been developed primarily for continuation and bifurcation detection in
boundary-value problems.

First, we convert (\ref{eq:sim}) to a system of first-order equations by introducing the state variables
\begin{equation}
\label{eq:simbcsu}
\!\!\!\!\!\!\!\!\!\!\!\!\!\!\!\!\!\!\!\!
u_1 = f, \,\,\, u_2 = f', \,\,\, u_3 = f'', \,\,\, u_4 = f^3\left(f''' + \frac{1}{\xi}f'' - \frac{1}{\xi^2}f'\right) + f^{2-n}f', \,\,\, u_5 = \xi.
\end{equation}
The state variables satisfy the system
\numparts
\label{eq:simodes}
\begin{eqnarray}
u_1' &= u_2, \\
u_2' &= u_3 , \\
u_3' &= \frac{u_4}{u_1^3} - \frac{u_3}{u_5} + \frac{u_2}{u_5^2} - \frac{u_2}{u_1^{n+1}}, \\
u_4' &= \alpha u_1 - \beta u_5 u_2 - \frac{u_4}{u_5}, \\
u_5' &= 1. \label{eq:simodesend}
\end{eqnarray}
\endnumparts
Note that $u_5$ has been introduced so that the system is autonomous.

As a starting solution, we note that when $n=3$, (\ref{eq:sim}) has the exact
solution $f_\textrm{e}(\xi) = c_0 \sqrt{\xi}$ satisfying the far-field
conditions (\ref{eq:simbc2}), but not conditions (\ref{eq:simbc1}) at
$\xi=0$. Equivalently, we have an exact solution $\bm u_\mathrm{e}(\xi)$ that
satisfies (\ref{eq:simodes}a--e) and far-field conditions
(\ref{eq:simbcsu}), but not the equivalent of the left-hand boundary
conditions (\ref{eq:simbc1}), namely $u_1 = u_4 = 0$ at $\xi=0$. We thus
introduce artificial parameters $\delta_1$ and $\delta_2$ into the left-hand
boundary conditions, as well as an approximate left-hand boundary location
$\xi_0 \ll 1$ to handle the coordinate singularity at $\xi=0$, and enforce
the conditions
\begin{equation}
u_1(\xi_0) = f_0, \qquad u_2(\xi_0) = \delta_1, \qquad u_4(\xi_0) = \delta_2.
\label{eq:simbc3}
\end{equation}
The far-field boundary conditions (\ref{eq:simbc2}) are also enforced at the large but finite value $\xi = L$.  Given appropriate values of $f_0$, $\delta_1$ and $\delta_2$, namely,
\begin{equation}
f_0 = c_0\xi_0^{1/2}, \quad \delta_1 = \frac{1}{2}c_0\xi_0^{-1/2}, \quad \delta_2 = \left(\frac{1}{2} - \frac{3}{8}c_0^4\right)\xi_0^{-1},
\end{equation}
$\bm u_\mathrm{e}(\xi)$ satisfies (\ref{eq:simbc3}), so may be used as a starting point for our computation.  Using numerical continuation, we now take $\delta_2$ and $\delta_1$ to zero for positive $\xi_0$, allowing $f_0$ to be free in each case.  Now, as $\xi_0$ is taken to zero, we approach a solution to the original problem (\ref{eq:sim}) satisfying the correct boundary conditions (\ref{eq:simbc1}).

The introduction of the artificial parameters also provides a systematic way of computing the other members of the discrete family of solutions. For $\delta_1=0$ and $\xi_0 > 0$, we allow $f_0$ to vary, letting $\delta_2$ be free.  The curve of $\delta_2$ against $f_0$ oscillates around $\delta_2=0$, each intersection corresponding to a solution of (\ref{eq:sim}).  This approach is similar to that used for the plane-symmetric problem \cite{Tseluiko2013}, although in our case the variation of the artificial parameters in the boundary conditions cannot take place for $\xi_0=0$ due to the coordinate singularity.  Instead, each solution starting from a zero of $\delta_2$ is continued in $\xi_0$, converging as $\xi_0 \rightarrow 0$.

Finally, after finding the discrete solutions for $n=3$, we continue in $n$ to trace out discrete solution branches.

\subsection{Results}

\begin{figure}
\centering
\includegraphics{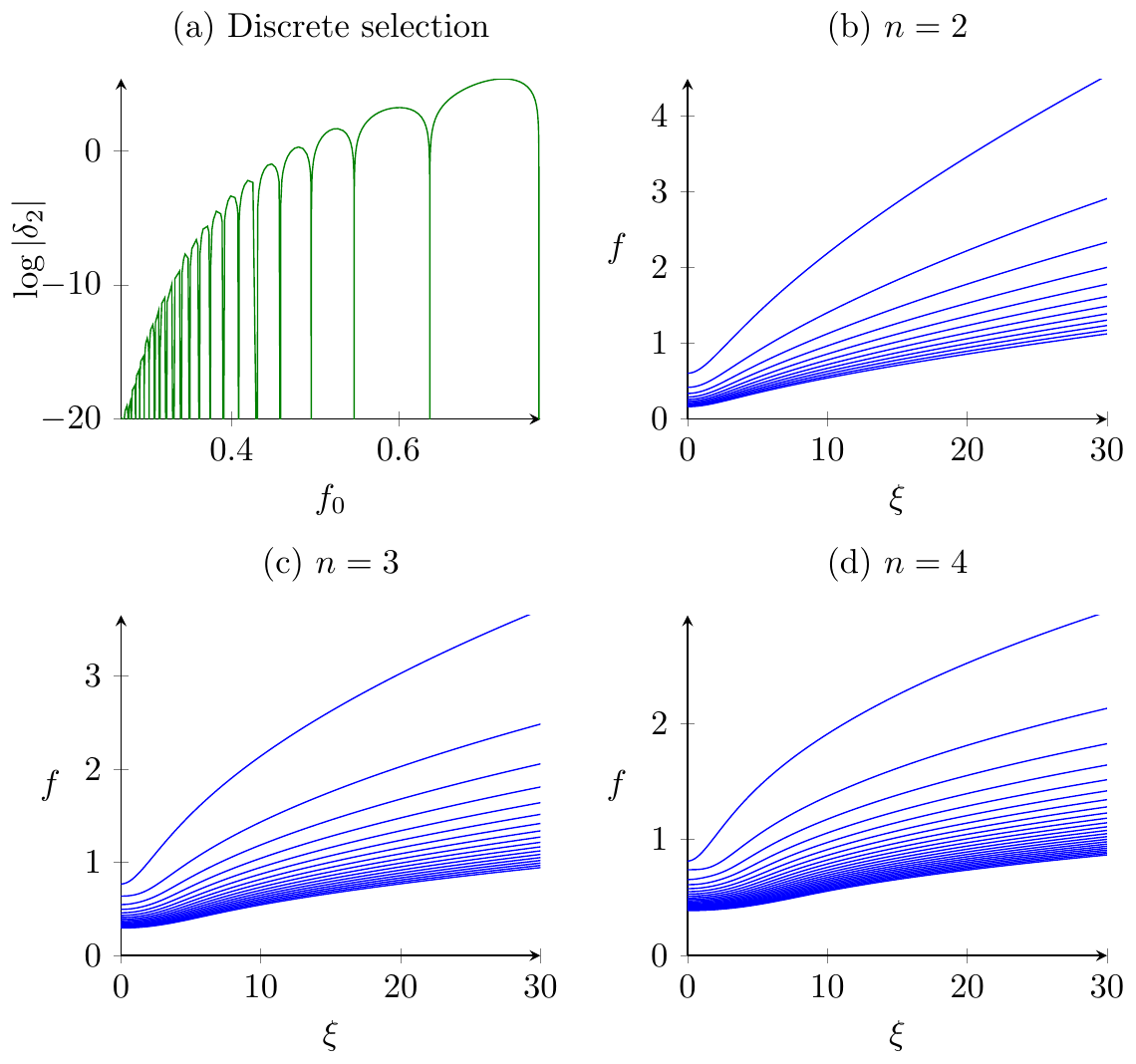}

\caption{(a) The artificial parameter $\delta_2$ as a function of the scaled film thickness at the origin $f_0$ for $n=3$.  The roots $\delta_2=0$ correspond to solutions of (\ref{eq:sim}). (b--d) The computed similarity solutions $f(\xi)$ for $n=2$, $3$ and $4$, respectively.}
\label{fig:sols}

\end{figure}

\begin{figure}
\centering

\includegraphics{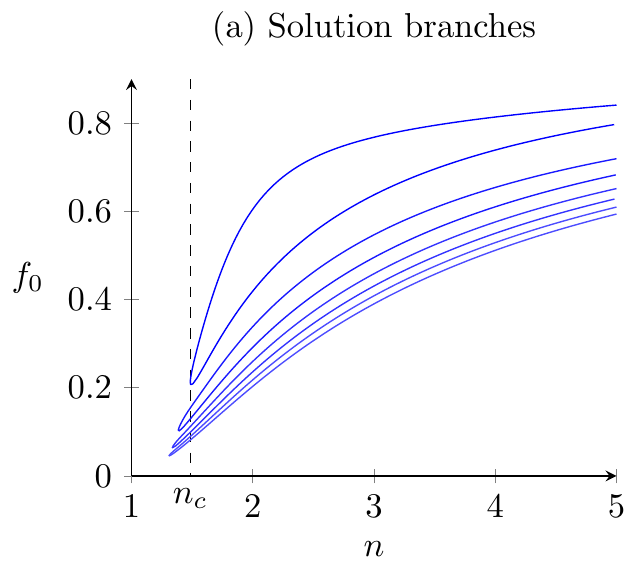}

\includegraphics{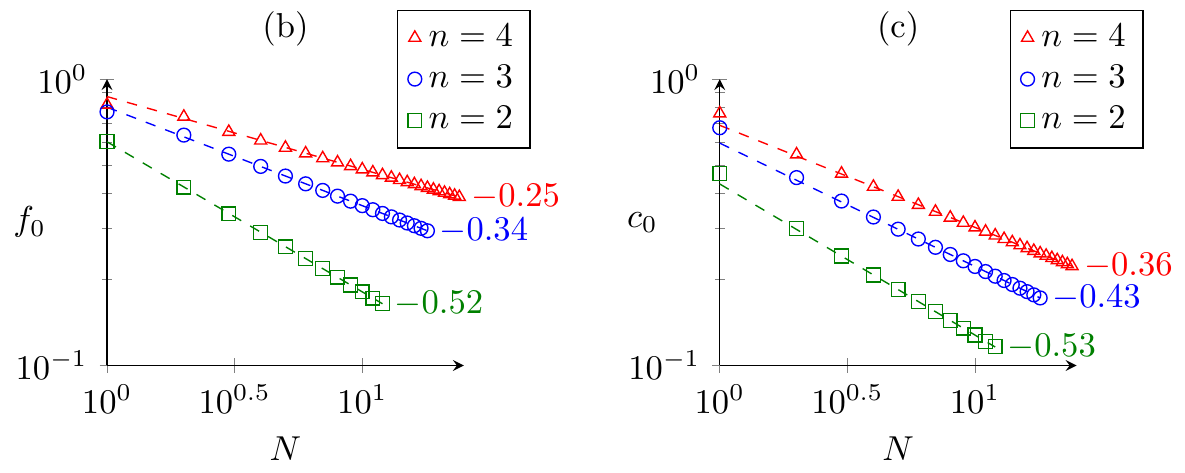}

\caption{(a) The first eight branches of solutions as a function of the disjoining pressure exponent $n$; each pair of branches merge at a turning point between $1$ and $2$.  The turning point in the first branch pair occurs at $n_c \approx  1.485$.
(b,c) The asymptotic power-law behaviour of the scaled thickness at the origin $f_0$, and the far-field coefficient ${c_0}$, for each of $n=2,3,4$, for large branch number $N$.  Power-law curves (dashed lines) are fitted to the last few points of the data (up to $N=12$ for $n=2$, $N=18$ for $n=3$ and $N=24$ for $n=4$).  The exponents are displayed next to each curve.
}
\label{fig:branches}
\end{figure}

In figure \ref{fig:sols}a, we plot the artificial parameter $\delta_2$ (using
a logarithmic scale) against $f_0$ for $n=3$, $\xi_0 = 10^{-4}$ and $L=100$, showing
the selection of discrete solutions corresponding to the roots $\delta_2=0$.
This figure is qualitatively similar to the corresponding figure for the
plane-symmetric case\,\cite[figure~5c]{Tseluiko2013}.

As the magnitude of oscillations of $\delta_2$ decreases, the selection of discrete solutions becomes more difficult to resolve.  This difficulty is more severe for smaller $n$.  We were able to reliably compute $24$ solutions for $n=4$,  $18$ solutions for $n=3$,  and $12$ solutions for $n=2$, out of a supposedly countably infinite set of such solutions.
In figures \ref{fig:sols}b--d, we plot the computed solutions for $n=2$,
$n=3$ and $n=4$, respectively.  Solutions with larger $n$ typically have larger values at the origin
$f_0$, and are shallower; as seen in (\ref{eq:alphabeta}), the far-field
exponent $\alpha/\beta = 2/(n+1)$ is decreasing in $n$.

In figure \ref{fig:branches}a, we plot the first eight discrete branches of
solutions, characterised by $f_0$, over a range of values of $n$. The most
interesting phenomenon to note is the merging of pairs of branches at a value
$n>1$ as $n$ decreases.  The branch with the largest value of $f_0$ merges
with the second branch at the critical value $n=n_c \approx 1.485$. At $n=3$,
this branch is the only stable one \,\cite{Witelski1999}, and it is
reasonable to assume these stability properties carry over to other values of
$n$. The stable solution branch then does not exist for $n < n_c$, and there
are therefore no stable similarity solutions in this region.

Subsequent pairs of solution branches also merge at turning points, as
depicted in figure \ref{fig:branches}a. While these turning points are less
than $n_c$, these solution branches are expected to be unstable.  Since we
compute eight solution branches in total, we have found four turning points
(including the first point at $n_c$).  The sensitivity of the numerical
problem for higher branches makes computation of further points highly
difficult, so we do not attempt to estimate the asymptotic behaviour of the
turning points (particularly given the unreliability of seemingly accurate
power-law fits as discussed below in section \ref{subsec:asymptotics}).
However, this is an interesting question that could possibly be answered
using asymptotic techniques, as discussed further in section
\ref{sec:conclusion}.

The non-existence of a stable similarity solution for $n < n_c$ raises the
issue of the dynamical behaviour of the time-dependent problem (\ref{eq:pde})
in this regime.  We explore the time-dependent behaviour in section
\ref{sec:numerics} via numerical computation.

\subsection{behaviour for high branch number}
\label{subsec:asymptotics}

In order to compare to and extend on previous results, we also compute the asymptotic behaviour of solutions for large branch number $N$.  In the literature, it is standard to characterise the solutions by the coefficient ${c_0}$ in the far-field condition (\ref{eq:ff}); we compute the dependence of both $f_0$ and ${c_0}$ on $N$.  Both $f_0$ and $c_0$ go to zero as $N\rightarrow\infty$, and both relationships are seemingly well fitted by power laws, with exponent dependent on $n$, as depicted in figures \ref{fig:branches}b and \ref{fig:branches}c, respectively.
Of particular note is the fact that the exponents for $f_0$ are roughy equal to $1/n$, while the exponent $-0.43$ for ${c_0}$ when $n=3$ is equal to the previously derived value \,\cite{Witelski2000}.

However, while the numerically computed results are seemingly well-fit by a
power law, the true asymptotic behaviour of these solution coefficients is
likely to be more subtle, as indicated by the recent study by Chapman et al.
\cite{Chapman2013}.  These authors performed the asymptotic analysis of the
plane-symmetric version of (\ref{eq:sim}) for $n=3$, using the techniques of
exponential asymptotics.  The main result is that the apparent power law
behaviour does not hold in the $N\rightarrow\infty$ limit; instead, they find that
the correct asymptotic behaviour is actually
\begin{equation}
\label{eq:asymp}
\frac{2}{5}{c_0}^{-2}\log({c_0}^{-1}) + a_1{c_0}^{-2} + a_2 \sim N, \qquad N \rightarrow\infty,
\end{equation}
for positive constants $a_1$ and $a_2$; these constants depend on the
leading-order (\mbox{large-$N$}) solution, which can only be solved numerically. A
similar formula to (\ref{eq:asymp}) is likely to be true for axisymmetric
rupture and general exponent $n$, so the numerically fitted power laws in
figure \ref{fig:branches} cannot be taken to be representative of the true
large-$N$ behaviour.

\subsection{Solution branches for the plane-symmetric problem}

Given the existence of branch-merging for axisymmetric thin films, it is worth checking to see whether the same phenomenon occurs in the plane-symmetric version of the same problem.  If the axisymmetric film ruptures in a self-similar way at a point away from the origin (corresponding to ring rupture), then the local behaviour is identical to plane-symmetric rupture \cite{Witelski2000}, so the existence of stable plane-symmetric solutions is therefore also of relevance to the fate of the time-dependent, axisymmetric film governed by (\ref{eq:pde}).

In a plane-symmetric geometry, the equation for the similarity solution $f$ is,
\begin{equation}
\label{eq:planesim}
-\alpha f + \beta\xi f' = -(f^3 f''' + f^{2-n}f')',
\end{equation}
where now $\xi = x/(t_0-t)^\beta$, with $x$ denoting the Cartesian coordinate.  Appropriate boundary conditions are (\ref{eq:simbc1}) at the origin, which are now interpreted as symmetry conditions, and the same stationarity conditions (\ref{eq:simbc2}) as for the axisymmetric problem.

The major difference in the plane-symmetric case is that, since the square
root function $f_\textrm{e}(\xi) = c_0\sqrt{\xi}$ is no longer an exact
solution, we must choose a different starting point.  For this reason, we
directly extend the solutions generated by \cite{Tseluiko2013} for $n=3$,
where the starting point for continuation was obtained by homotopy
continuation from the marginal stability problem of a near-flat film, to the
similarity equation (\ref{eq:planesim}). Another difference is that the
boundary conditions may be imposed at $\xi=0$ without encountering a
coordinate singularity. Otherwise, the process of computing solution branches
over $n$ is the same as carried out above for the axisymmetric problem.

\begin{figure}
\centering
\includegraphics{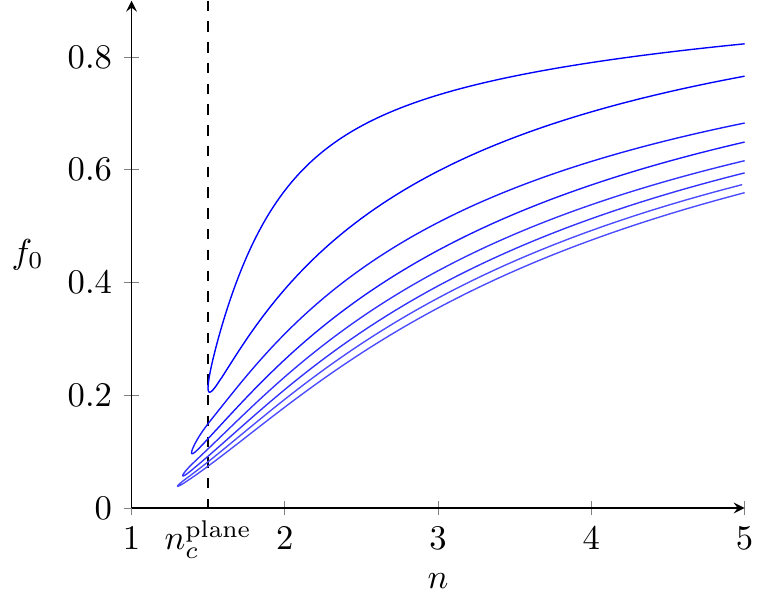}

\caption{Solution branches (disjoining pressure exponent $n$ against $f_0$) of the self-similar rupture problem with plane symmetry (\ref{eq:planesim}).  The first turning point $n_c^\mathrm{plane} \approx 1.50$ (to two decimal places) is slightly above the first turning point $n_c$ in the axisymmetric problem.  The following three turning points are at $1.39$, $1.33$, and $1.30$.}
\label{fig:planesim}
\end{figure}

The solution branches thus computed are plotted in figure \ref{fig:planesim}.
It is clear that the different geometry has only a minor effect on the
branches, and branch pairs still merge as $n$ decreases.  In fact, the first
turning point, which we denote by $n_c^\mathrm{plane}$, is equal to $1.5$ within
numerical error, which is slightly larger than the corresponding value $n_c
\approx 1.485$ in the axisymmetric case.  Thus, when $n < n_c <
n_c^\mathrm{plane}$, there are no stable similarity solutions of either plane
or axisymmetric form.

\section{Time-dependent computation}
\label{sec:numerics}
\subsection{Instability of a flat interface}

We now compute solutions to the time-dependent evolution of a rupturing film
(\ref{eq:pde}), and we contrast our results with the similarity solution for
different disjoining pressure exponents $n$.  In particular, we focus on the
difference in behaviour between $n > n_c$, where a stable axisymmetric
similarity solution exists, to $n < n_c$, where no stable axisymmetric
similarity solution exists.

The time-dependent computations are performed on a finite spatial domain $0 <
r < R$, such as depicted in figure \ref{fig:schematic}. We thus need two
boundary conditions at $r=0$ and two conditions at $r=R$. These are
\begin{equation}
\label{eq:bcs}
h_r = 0, \qquad rq \rightarrow 0 \qquad \textrm{at $r=0$ and at $r=R$}
\end{equation}
Recall that $q$ is the flow rate in the radial direction per unit length in the
angular direction, as defined in (\ref{eq:pde}).  The conditions at $r=0$ are necessary for smoothness of the solution at
$r=0$.  By expanding the expression for $q$ near $r=0$, it is easily obtained
that the zero-flux condition at $r=0$ is equivalent to $h_{rrr}=0$. The zero-flux condition at $r=R$ fixes the amount of mass in the system, while the
zero-slope condition is chosen for convenience; with such a condition a
uniform film $h \equiv 1$ is an exact solution, whose linear stability can be
determined analytically.  A more realistic condition may involve a prescribed
contact angle $h_r = \alpha$, or a pinned contact line $h = h_R$, but in any
case the condition should not affect self-similar rupture at $r=0$, so is not
important for our purposes.

Recall that  for the chosen non-dimensionalisation of (\ref{eq:pde}), the dimensionless average film thickness is unity, while the domain size $R$ is a parameter of the problem.
If the derivative of disjoining pressure is such that $\Pi'(1) > 0$, the flat
film $h\equiv 1$ is unstable to perturbations of sufficiently long
wavelength. There is, therefore, a critical domain size $R$ above which the
flat film becomes unstable, as derived by \cite{Witelski2000}; since the
linearisation of (\ref{eq:pde}) with conditions (\ref{eq:bcs}) is
self-adjoint, the critical value of $R$ may be determined from the marginal
stability problem:
\begin{equation}
\label{eq:stab}
\tilde h'' + \frac{1}{r}\tilde h' + \Pi'(1) \tilde h = 0, \qquad  \tilde h'(0) = \tilde h'(R) = 0.
\end{equation}
Here, $R$ is an eigenvalue.  The solution to (\ref{eq:stab}) may be
represented exactly in terms of zeroth-order Bessel functions of the first
kind, and subsequently the critical value $R = R_c$ is given by the smallest
root of
\begin{equation}
\label{eq:Rc}
J_1(\sqrt{\Pi'(1)}R) = 0 \qquad \Rightarrow \qquad R_c \approx \frac{3.83}{\sqrt{\Pi'(1)}},
\end{equation}
where $J_1$ is the first-order Bessel function of the first kind. By our
choice of scaling to reach the non-dimensional equation (\ref{eq:pde}), $\Pi'(1) = 1$ for all $n>0$.  We
must therefore choose a domain size $R$ greater than $R_c\approx 3.83$ in
order to observe instability leading to rupture.  In fact, for $n=3$,
Witelski and Bernoff \cite{Witelski2000} showed that there exists a small
interval $R \in (R_c, R_c+\epsilon)$ in which there is a stable,
axisymmetric, non-uniform solution.  In our computations below, we adopt
$R=4$, which is sufficiently large so that the uniform film is seen to be
unstable and a non-uniform stable solution does not exist in all the
computations we carry out.

\subsection{Method}

The solution is obtained using the finite-volume method\,\cite{Patankar1980}.
This method involves explicit computation of the fluxes, which means that
conservation of mass is guaranteed and flux conditions are easy to implement
on the boundaries. The domain $r \in [0,R]$ is divided into $M$ cells or
`control volumes' by faces at $r = r_{\textrm{f}, i}, \ i = 1,\,\ldots,\, M+1$,
with $r_{\textrm{f},1} = 0$ and $r_{\textrm{f}, M+1} = R$.  The value of $h$ is
approximated by its value in the center of each cell $r_i$: $h_i = h(r_i), \
i = 1,\,\ldots,\, M$.

The flux $q_{\textrm{f}, i}$ at each interior face is computed from the expression for $q$ in (\ref{eq:pde}) by averaging for $h$ , the two-point central finite-difference approximation for $h_r$, and four-point approximations for $ h_{rr}$ and $h_{rrr}$. The zero-slope boundary conditions are built into the central-difference approximations for the derivatives of $h$ at the first and last interior faces, while the zero-flux conditions give the flux for the outer faces.  The change in thickness then comes from conservation of mass:
\[
\dot h_{i} = -\frac{r_{\textrm{f}, i+1}\,q_{\textrm{f}, i+1} - r_{\textrm{f}, i}\,q_{\textrm{f}, i}}{r_i\, \delta r_i}, \qquad \delta r_i = r_{\textrm{f}, i+1} - r_{\textrm{f}, i}.
\]
The values of $h$ are advanced in time using MATLAB's \texttt{ode15s} algorithm.

\subsection{Results}

\begin{figure}
\centering
\includegraphics{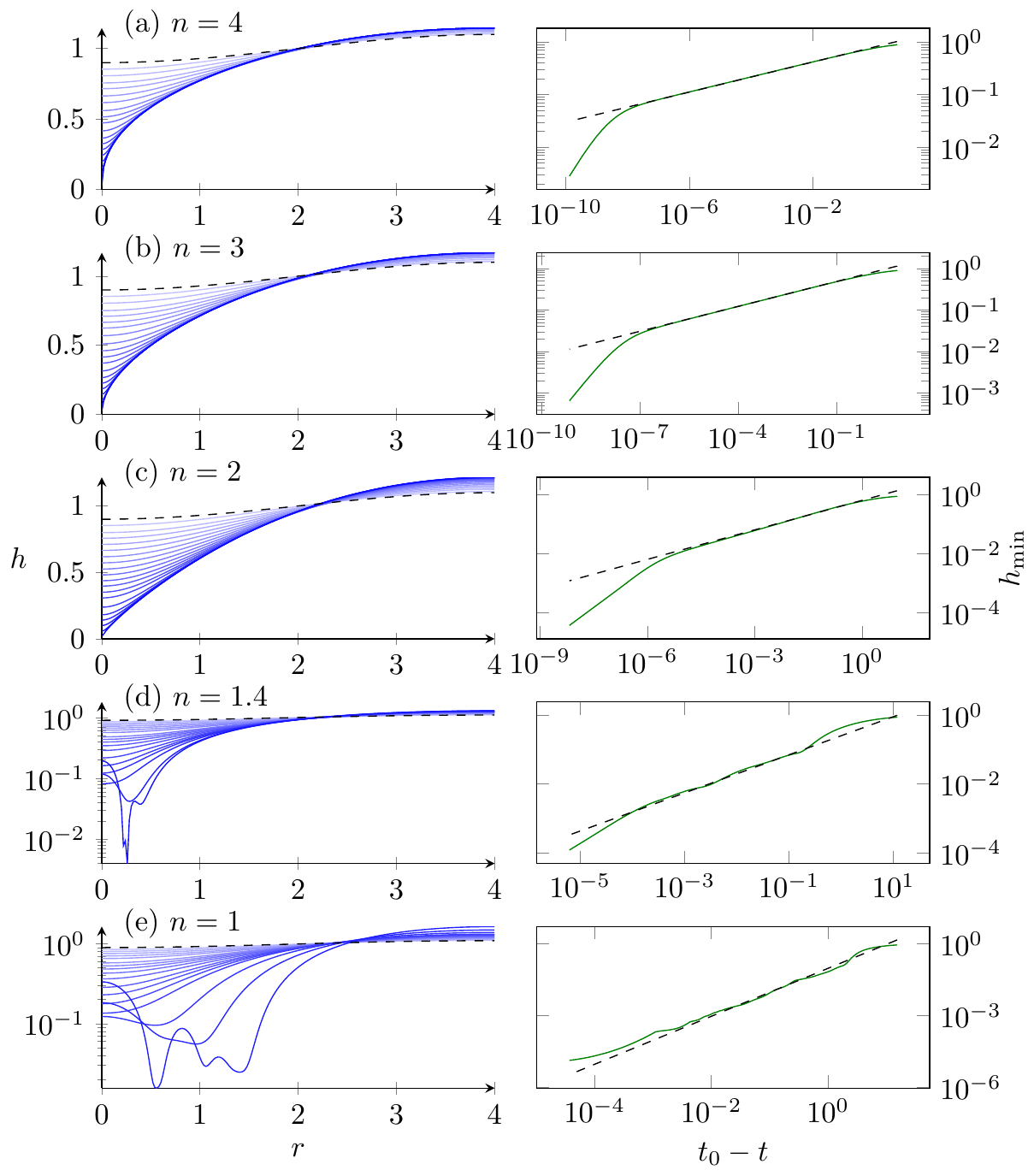}

\caption{Time-dependent solutions to the boundary-value problem (\ref{eq:pde}), (\ref{eq:bcs}) with disjoining pressure exponents $n$ given by: (a) $n=4$, (b) $n=3$, (c) $n=2$, (d) $n=1.4$, and (e) $n=1$.
When $n$ is greater than the critical value $n_c$ (a--c), an initially perturbed flat film (dashed line) tends to point rupture at the origin $r=0$ (line shading is darker in increasing time; profiles are not plotted at equally spaced time intervals).  The numerical behaviour of $h_\mathrm{min}(t) = \min_{0\leq r< \leq R} h(r,t)$ is also depicted in the right-hand panels.   When $n > n_\mathrm{c}$, $h_\mathrm{min}$ behaves as a power law in $t_0-t$, with exponent $\alpha$ given by (\ref{eq:alphabeta}) (the theoretical slope is indicated by the dashed line), as expected for self-similarity (departure from this power law for very small $h_\mathrm{min}$ is due to discretisation error).
When $n < n_\mathrm{c}$ (d--e), the film does not rupture at the origin, instead destabilising into a cascade of oscillations of decreasing wavelength.  The minimum thickness $h_\mathrm{min}$ only approximately follows a power law with appropriate values of $\alpha$.
}
\label{fig:numsols1}
\end{figure}

\begin{figure}
\centering
\includegraphics{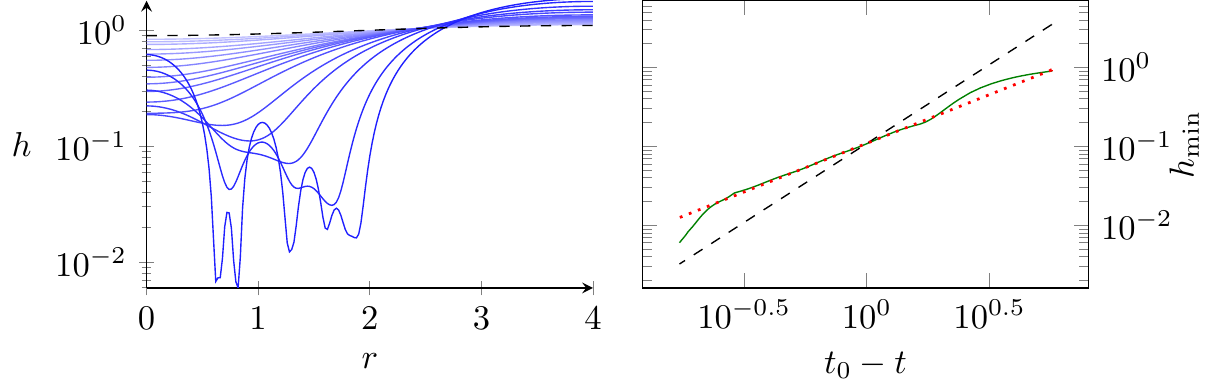}

\caption{Time-dependent solutions to the boundary-value problem (\ref{eq:pde}), (\ref{eq:bcs}) with disjoining pressure exponent $n = 0.75$ (line shading is darker in increasing time).  The film exhibits cascading oscillations, while the minimum thickness approximately follows a power law.  Unlike the cases for $n>1$ (fig.~\ref{fig:numsols1}), however, the power law is not that predicted by the self-similar ansatz $\alpha = 2$ (black dashed line); instead, fitting a power law model to the numerical results (red dotted line) gives an exponent approximately equal to $1.2$.}
\label{fig:numsols1a}
\end{figure}

\begin{figure}

\centering

\includegraphics{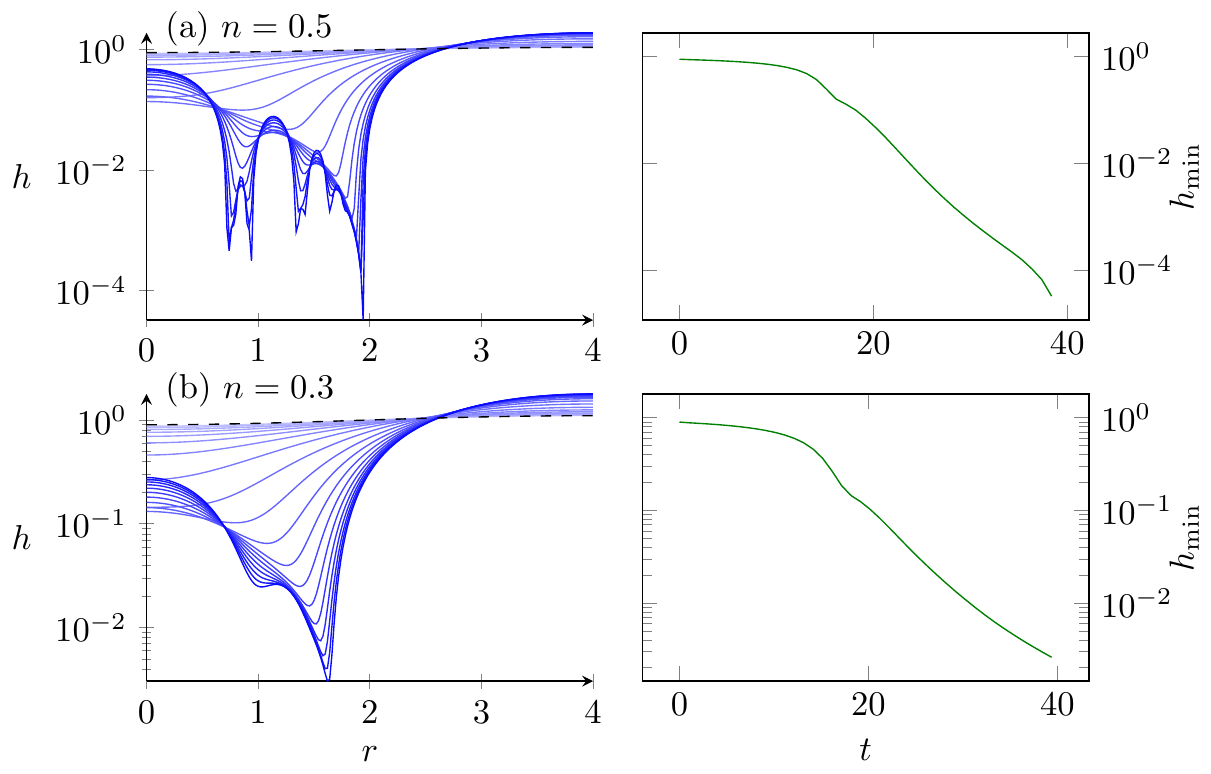}

\caption{Time-dependent solutions to the boundary-value problem (\ref{eq:pde}), (\ref{eq:bcs}) with disjoining pressure exponents (a) $n=0.5$, and (b) $n=0.3$ (line shading is darker in increasing time).  At these values, $h_\mathrm{min}$ decays exponentially, and may not rupture in finite time at all.}
\label{fig:numsols2}
\end{figure}

We present the results of computations for a range of disjoining pressure
exponents $n$, in figure~\ref{fig:numsols1} ($n \geq 1$), figure~\ref{fig:numsols1a} ($n=0.75\in(1/2,1)$) and
figure~\ref{fig:numsols2} ($n \leq 1/2$).  As previously stated, the domain size
is $R=4$. In each case, the initial condition is a uniform film perturbed by
a small cosine perturbation, $h(r,0) = 1 - 0.1\cos(\pi r/R)$, and $M=2000$ nodes
are used.  For $n < 1/2$, the computations are terminated when the minimum
thickness is less than a threshold value of $10^{-5}$.

In figure \ref{fig:numsols1}, the film evolves to the predicted self-similar behaviour, rupturing at the point $r=0$ at a finite time $t_0$.  As well as plotting the solution profiles we also observe self-similarity by computing the behaviour of the minimum thickness $h_\mathrm{min}(t) = \min_{0\leq r \leq R} h(r,t)$.  For the values of $n > n_c$, the minimum always occurs at $r=0$ and behaves proportional to $(t_0-t)^\alpha$, where $\alpha$ is as in (\ref{eq:alphabeta}) for the given $n$.

On the other hand, when $n < n_\mathrm c$, solutions are not attracted to similarity solutions exhibiting point rupture.  Instead, as the film thins, it becomes unstable to short wavelength perturbations and develops a cascade of oscillatory humps.  Nevertheless, it is still of interest to observe the behaviour of the minimum film thickness $h_\mathrm{min}$ over time.  For the values $n=1.4$ and $n=1$ depicted in figures \ref{fig:numsols1}d and e, respectively, $h_\mathrm{min}$ is monotonically decreasing but exhibits features that clearly correspond to the change in position at which $h$ attains its minimum.  Despite these features, $h_\mathrm{min}$ still behaves roughly in accordance to the power law predicted from the assumption of self-similarity (\ref{eq:alphabeta}), and appears to rupture at a point.

As noted in section \ref{sec:similarity}, when $1/2 < n < 1$, the similarity ansatz is still formally valid, but the far-field quasi-stationary condition (\ref{eq:ff}) is no longer applicable.  The results for $n=0.75$, included in figure \ref{fig:numsols1a}, show that the thickness profile develops oscillatory structures and $h_\mathrm{min}$ behaves approximately as a power law, but not of the value $\alpha=2$ predicted by the self-similar ansatz (\ref{eq:alphabeta}).  Instead, by numerically fitting to a large subset of the data we find that the power law exponent is approximately $1.2$.  This behaviour cannot therefore be explained by the similarity ansatz (\ref{eq:similarity_ansatz}).  A similar power law exponent is observed for other values of $n \in (1/2,1)$.

Also as described in section \ref{sec:similarity}, the value $n = 1/2$ is critical, as at this value the similarity exponents $\alpha$ and $\beta$ in (\ref{eq:alphabeta}) are undefined; instead, the appropriate similarity ansatz uses exponential, rather than power-law, scaling.  When $n < 1/2$, the corresponding values of $\alpha$ and $\beta$ are both negative, and the assumptions made in forming the similarity ansatz are no longer valid.   In order to gain insight into this regime, we also perform computations for $n = 1/2$ and $n = 0.3 < 1/2$. The results are plotted in figure \ref{fig:numsols2}, along with the evolution of the minimum film thickness $h_\mathrm{min}(t)$ over time.  In this regime $h_\mathrm{min}$ decays in a roughly exponential fashion.  Again we see the formation of a cascade of oscillatory structures.

 It is interesting to note that this secondary behaviour has been observed in previous numerical results pertaining to a thin film destabilised by thermocapillary Marangoni stress, not only in the lubrication approximation\,\cite{Joo1991} but also using the full Stokes\,\cite{Boos1999} and Navier-Stokes\,\cite{Krishnamoorthy1995} equations.  As mentioned in the introduction, the destabilising effect of thermocapillarity is equivalent to a logarithmic disjoining pressure, and corresponds to the special case $n=0$ in our formulation.  Our results show that this qualitative behaviour extends to higher values of $n$.

For $n > n_c$, the existence of stable self-similar solutions means that finite-time film rupture is generic in this regime.  What is much more difficult to ascertain from the numerics is whether the film ruptures  in finite time for smaller values of $n$.  The behaviour of $h_\textrm{min}$ suggests that for $1/2 < n < n_\mathrm c$, the minimum thickness behaves roughly as a power law and thus  tends to zero in finite time, while for $n \leq 1/2$ it decays exponentially to zero as $t\rightarrow \infty$.  However, the numerical scheme cannot reliably capture the behaviour of solutions of thickness less than the spatial discretisation, a problem encountered by other studies of this problem \cite{Boos1999,Joo1991,Krishnamoorthy1995}.  There is therefore scope for analytical results in this area, as we discuss in the conclusion.

\vspace{0.4cm}
\section{Conclusion}
\label{sec:conclusion}

In this study, we have determined self-similar, axisymmetric rupture
solutions for a generic thin-film equation that arises in the study of film
dewetting due to intermolecular forces. We have shown that for sufficiently
large disjoining pressure exponent $n$, discrete solutions exist.  Pairs of
discrete solution branches merge as $n$ decreases, leading to a critical
value $n_c \approx 1.485$ below which stable self-similar point rupture does
not occur. We also numerically computed solution branches for the
plane-symmetric version, showing that the first pair of solution branches
merges at $n_c^\mathrm{plane} \approx 1.5$.  Thus, for $n < n_c$, neither
self-similar point nor line rupture (nor ring-rupture in an axisymmetric
setting, which behaves locally like line rupture) is stable, and films will
evolve in a non-self-similar fashion. This prediction is supported by the
numerical solution of the full time-dependent problem.

The numerical results also suggest that when $n \leq 1/2$, finite-time
rupture may not occur at all. It is certainly plausible that there exists a
critical value $n = n^* < n_c$ below which finite-time rupture cannot happen,
even in a non-self-similar fashion.  There are some related examples for
other equations of the form (\ref{eq:genpde}). For instance, Bertozzi et
al.~\cite{Bertozzi1994} explore the possibility of rupture for the general
equation (\ref{eq:genpde}) with $\mathcal M = h^p$ and $\mathcal N = 0$.  In
that case the possibility of finite-time rupture depends on the exponent $p$,
and it is proven that finite-time rupture cannot occur when $p
> 5/2$.  On the other hand, if $\mathcal N = -h^q$ then the possibility of
finite-time blow-up ($h\rightarrow\infty$ as $t\rightarrow t_0$) is strongly
dependent on the exponents $p$ and $q$\,\cite{Bertozzi1998}. While our
numerical results are not conclusive, they do point to an interesting
conjecture, that finite-time rupture for (\ref{eq:pde}) does not occur for $n
\leq 1/2$. It is worth further study to see if such a result can be
established rigorously, perhaps using similar methods as those in
\cite{Bertozzi1994}.

Multiple discrete solutions were found via homotopy continuation by allowing
an artificial error parameter $\delta_2$ in the left-hand boundary conditions to
vary.  This method is similar to the one used previously in the plane
symmetric case\,\cite{Tseluiko2013}.
However, the axisymmetric case is made more complicated by the fact that the
left-hand boundary condition must be applied at a point $\xi_0 > 0$, in order
to avoid the coordinate singularity at $\xi=0$. An alternative approach would
be to introduce an artificial parameter into the system (\ref{eq:simodes}) in
such a way that the boundary conditions are maintained, and the artificial
parameter oscillates around zero as the solution value at the origin, $f_0$,
is decreased.  Experimentation with the modified system, however, did not
result in any obvious way of introducing such a parameter, so we leave this
as a topic for future research.

The presence of cascading oscillations in the parameter regime $n <
n_\mathrm{c}$ may indicate the existence of `limit-cycle' solutions in the
self-similar coordinate system, where the steady states (that is, the
self-similar solutions) are unstable.  The phenomenon of limit cycles in
self-similar coordinates has been referred to as `discrete self-similarity',
as the self-similarity is seen periodically in (logarithmic)
time\,\cite{Eggers2015}.  Computation of these limit-cycle solutions will
require careful linear stability analysis of the higher solution branches (of
the plane-symmetric problem, as for $n < n_c$ rupture does not generically
occur at the origin) to detect any Hopf bifurcations. In addition, it would
be very useful to perform time-dependent computations in the self-similar
coordinates (with a logarithmic time variable); however, this is difficult in
practice due to exponentially growing modes that correspond to translations
in space and time~\cite{Eggers2015,Witelski1999,Witelski2000}, and is
therefore left as a topic for future investigation.

Finally, we note that the exponential asymptotic analysis of the plane-symmetric
rupture problem with $n=3$ could be extended to both the axisymmetric
geometry and general $n$, likely leading to asymptotic behaviour such as
(\ref{eq:asymp}) for the far-field coefficient ${c_0}$ (appearing in
(\ref{eq:ff})) as the branch number $N$ becomes large.  Whether it will have
the same form as (\ref{eq:asymp}) with different coefficients, or different
powers on ${c_0}$, is not immediately clear; in any case, as this study (and
previous power-law estimates in \cite{Tseluiko2013,Witelski1999}) indicates,
obtaining such a delicate asymptotic behaviour from numerical results alone is
practically impossible, and using techniques from exponential asymptotics is
required. It is also likely that the asymptotic position of the turning
points for large branch number may be obtained from such an analysis.  It is
open as to what the asymptotic behaviour of these turning points is; if, as we
suggested above, there is a critical value $n = n^*$ below which no
finite-time rupture is possible, this may coincide with the limiting behaviour
of the turning points in the solution branches, as in this case we would not
expect to have even unstable similarity solutions for $n < n^*$. We shall
consider this and related issues in future studies.

\section*{Acknowledgments}
This study is a byproduct of extensive discussions with Howard Stone on
thin-film rupture. We are grateful to him for his role in originating our
collaboration on this problem and for insightful comments and to Edgar
Knobloch for numerous stimulating discussions. We acknowledge financial
support from the Engineering and Physical Sciences Research Council (EPSRC)
of the UK through Grants No. EP/K008595/1 and EP/L020564/1. The work of DT
was partly supported by the EPSRC through Grant No. EP/K041134/1.


\end{document}